# Smartphone microendoscopy for high resolution fluorescence imaging

Xiangqian Hong,[a,*] Vivek K. Nagarajan,[a] Dale H. Mugler,[a] and Bing Yu[a]

[a]University of Akron, Department of Biomedical Engineering, Akron, OH, 44325, USA

**Abstract**. High resolution optical endoscopes are increasingly used in diagnosis of various medical conditions of internal organs, such as the gastrointestinal tracts, but they are too expensive for use in resource-poor settings. On the other hand, smartphones with high resolution cameras and Internet access have become more affordable, enabling them to diffuse into most rural areas and developing countries in the past decade. In this letter we describe a smartphone microendoscope that can take fluorescence images with a spatial resolution of 3.1 µm. Images collected from *ex vivo*, *in vitro* and *in vivo* samples using the device are also presented. The compact and cost-effective smartphone microendoscope may be envisaged as a powerful tool for detecting pre-cancerous lesions of internal organs in low and middle income countries.

Keywords: smartphone, optical endoscopy, fluorescence imaging.

Billions of people worldwide live in low and middle income countries (LMIC) where incidence and mortality rates of many medical conditions, such as oral, cervical and gastrointestinal (GI) cancers, are disproportionately high and adverse.[1] Lack of access to imaging equipment, such as high resolution optical endoscopes, and well-trained medical staffs are among the major factors responsible for the delayed diagnosis and high death rates in LMIC. There is an urgent need of affordable and easy-to-use high resolution endoscopes to improve the screening and early diagnostic rates of many medical conditions in low-resource settings.

Optical endoscopy is a powerful tool for noninvasive imaging of hollow tissue cavities through a catheter or minimally invasive imaging deep within tissue through a needle or laparoscopic/robotic instrument.[2] Various modern imaging modalities, including confocal microscopy,[3] fluorescence imaging,[4,5] optical coherent tomography (OCT),[6] photoacoustic imaging (PAI),[7] with cellular to subcellular resolution have been successfully incorporated into endoscopes. These emerging high resolution endoscopes show great potential in improving the accuracy for disease diagnosis, such as early cancer detection.[5,8,9] Most optical endoscopes employ an optical fiber, fiber optic imaging bundle, or light-guide for light delivery and collection. However, such high resolution endoscopic systems usually consist of bulky, power-consuming and expensive optical components, including thermal lamps, cooled cameras, discrete lens and filters, and/or galvanometer scanners as well as a computer, which make them unsuitable for applications in LMIC.

On the other hand, the cost of wireless technology has decreased over the years, making smartphones, a subset of mobile phone, a very affordable device, even for people living in many rural areas of developing countries. There are 7 billion mobile phone and 2.3 billion mobile-broadband subscriptions in 2014, with over 77% and 55%, respectively, in LMIC.[10] In addition to high resolution cameras, smartphones also offer enormous computation power, Internet access and other sensors on a compact platform. Smartphone-based diagnosis promises to reduce healthcare costs and provide access to advanced laboratories and experienced physicians in developed areas, thus revolutionizing healthcare in LMIC.

Smartphones are playing an emerging role in optical imaging for medical and biological applications. Breslauer *et al.*[11] reported a mobile phone microscope with a field-of-view (FOV) of 180 µm in diameter and a 1.2 µm resolution for the diagnosis of hematologic and infectious diseases. Switz *et al.*[12] added a reversed camera lens to a mobile phone to enable high-quality imaging over a FOV of ~10 mm$^2$ and successfully identified red and white blood cells in blood smears and soil-transmitted helminth eggs in stool samples. Tseng *et al.*[13] demonstrated a lens-free holographic microscope on a mobile phone that has been used to image various sized microparticles. Zhu *et al.*[14] reported wide-field FLI on a smartphone over a FOV of ~81 mm$^2$ with a resolution of ~20 µm. Most smartphone imaging devices utilize an external attachment to the rear camera of a mobile phone, but only a few have been designed for a non-fiber-optic endoscope. Wu *et al.*[15] transformed a smartphone into an endoscope for acquiring otorhinoscopic images from six patients for remote diagnosis. Jongsma *et al.*[16] developed a mobile phone otoendoscope, which has been commercialized by Endoscope-i Ltd. MobileODT has recently marketed a multimodal smartphone imaging system for cervical cancer detection.[17]

In this letter we present the design of a smartphone-based fiber optic microendoscope for high resolution fluorescence imaging and some preliminary experimental results. A schematic diagram of the smartphone microendoscope and an experimental setup to test the feasibility of the design are shown in Fig. 1. The system consists of a smartphone with a rear camera, imaging optics which will be engineered as an attachment in the future, and a fiber optic imaging bundle. The imaging optics includes a blue LED with a condenser lens (L1) and a band-pass filter for fluorescence excitation (BP1), a dichroic beamsplitter

*Address all correspondence to: Xiangqian Hong, E-mail: xh13@zips.uakron.edu

(DBS), a finite microscope objective (OBJ), a band-pass fluorescence emission filter (BP2), an eyepiece (EP), a FC/PC fiber optic connector and batteries. The filtered excitation light is redirected by the beamsplitter towards the objective to achieve a Kohler illumination (uniform illumination) on the proximal end of the fiber bundle plugged into the FC/PC connector. The distal end of the imaging bundle is in contact with the target being tested, such as a biological tissue. The fluorescence emissions from the target are collected by the same fiber bundle, propagate through the objective, beamsplitter and emission filter that blocks the excitation lights, and then enter the rear camera of the smartphone after being collimated by the eyepiece. The fluorescence image can be processed by the smartphone or wirelessly transmitted to a remote computer for analysis.

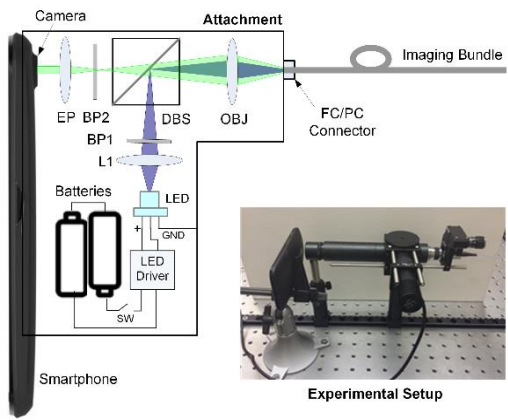

**Fig. 1** (a) Schematic diagram and (b) photograph of the experimental setup of the smartphone-based fiber optic microendoscope. LED – light emitting diode, GND – ground, L1 – lens 1, BP – band-pass filter, DBS – diachronic beamsplitter, OBJ – objective lens, and EP – eyepiece.

The choice of the LED (455 nm, M455L2, Thorlabs), exitation filter (FF01-452/45, Semrock), dichroic beamsplitter (495 nm cutoff, 475DCXRU, Chroma Technology) and emission filter (FF01-550/88, Semrock) is based on the use of proflavine as the fluorescence contrast agent. Proflavine, a topically applied DNA dye, has been previously used by Quinn and Muldoon et al to image cell nuclei for neoplasia detection in the cervix, oral cavity and Barrett's esophagus.[5,18,19] It has a peak excitation and emission wavelength of 445 nm and 515 nm, respectively. The fiber bundle (FIGH-30-650S, Fujikura) has an imaging area of 600 μm and consists of ~30,000 individual fibers of ~2 μm in diameter with a center-to-center distance about 3 μm. The rear camera of the Motorola smartphone Moto G has a 2592×1944 pixels at a size of 1.4 μm.

In the experimental setup shown in Fig. 1, a 20× finite objective and a 16× wide-field eyepiece were selected in combination with the cellphone camera lens to obtain a proper magnification. The actual imaging area filled 1730 pixels in diameter of the camera sensor array, which represents an image size of 1730×1.4 μm ≈ 2.4 mm in diameter, resulting in 4× magnification (2.4 mm/0.6 mm). Therefore, each individual fiber occupied about ~36 pixels of a raw image. The locations of the proximal end of the fiber bundle and the condenser were adjusted so that all pixels in the bundle were uniformly illuminated and clearly imaged onto the camera. A green fluorescence reference slide (2273-G, Ted Pella) was used to check the uniformity of the system. Fig. 2(a) shows a representative image of the proximal end of the fiber bundle when its distal end was in contact with the reference slide. A close look of the image marked by the red box in Fig. 2(a) indicates that individual fibers of the bundle were well resolved.

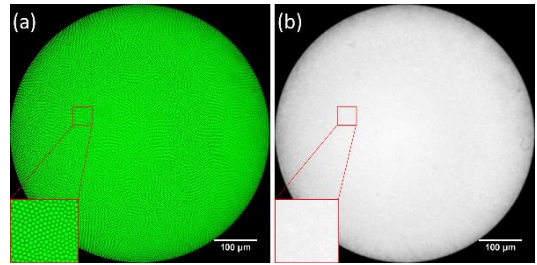

**Fig. 2** (a) Raw and (b) fiber pattern rejected images of a uniform fluorescence reference slide.

The raw image not only contains the structural information of the imaged target, but also carries the honeycomb fiber patterns. We employed the method proposed by Elter et al.[20] to eliminate the fiber pattern artifacts in the fiber bundle imaging. First, the imaging area was defined as the region of interest (ROI) and circularly cropped out of the image. Since the intensity within each individual fiber has a Gaussian distribution, the intensity at the center of each fiber represents the fluorescence intensity collected by the fiber. The built-in Matlab function 'imregionalmax' was then applied to locate the fiber centers and extract their intensity values. Because the 'imregionalmax' function can only process images with regional peaks that have a maximum connectivity of 26 or less, the raw images with a fiber occupying 36 pixels was scaled to half of its original size. Given that the resolution of the imaging bundle is limited by the center-to-center distance of two adjacent fibers, reducing the image size to half doesn't change the spatial resolution of the system. Next, the image was converted to a gray-scale intensity image. The final step was to assign the extracted center pixel values to the neighboring pixels to construct a comb structure free image. The fiber pattern rejected image of Fig. 2(a) is presented in Fig. 2(b). The pixelation artifacts were effectively removed in the reconstructed image.

To characterize the spatial resolution of the smartphone microendoscope, fluorescence images were taken from a 1951 USAF resolution test target that was placed on top of a green fluorescent reference slide. Fig. 3 (a) and (d) show the raw and fiber pattern rejected images of the test target, respectively. The intensity function across the lines (not shown) indicates that the valley intensity between the Group 7 Element 2 lines is 3dB below the peak value, while less than 3dB for that of the Group 7 Element 3 lines. This demonstrates that the microendoscope successfully resolved the adjacent lines of Element 2 in Group 7, as can aslo be visually seen from the enlarged area in Fig. 3(d). Thus, the spatial resolution was estimated to be about 3.1 μm. This value meets our expectation that the resolution of the setup is limited by the center-to-center distance between two adjacent fibers of the imaging bundle, which is about 3 μm.

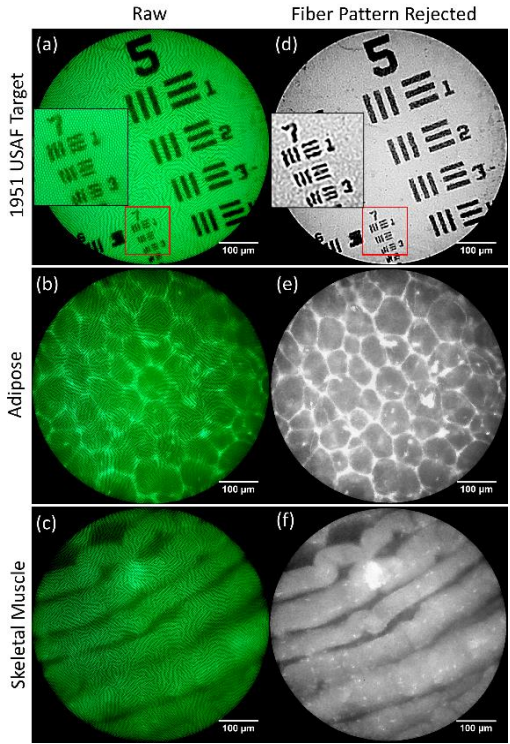

**Fig. 3** Raw and fiber pattern rejected fluorescence images taken from a 1951 USAF resolution target (a, d), an *ex vivo* porcine adipose tissue (b, e) and *ex vivo* bovine skeletal muscles (c, f).

*Ex vivo* porcine adipose and bovine skeletal muscle tissues were also imaged using the experimental setup in Fig. 1. Fresh porcine and bovine tissues were obtained from a local butcher's shop within 3 hours of the slaughter of the animals. Experiments were conducted immediately after the tissues were transported to the lab in a cooler. The tissues were sliced into a dimension of 2×2×1 cm (W×D×H). Proflavine at a concentration of 0.01% wt/vol (in PBS) was applied on the surface of the sliced tissues using a cotton swab and fluorescence images were taken immediately after in a dark room. The typical images are presented in Fig. 3(b) and 3(e) for the adipose tissue and Fig. 3(c) and 3(f) for the skeletal muscle. The white fat cells and muscle fascicles are both clearly visible. The brighter backgrounds between the cells of the adipose samples are likely due to the non-specific binding of excessive proflavine on the tissue which may be reduced by rinsing the sample before imaging.

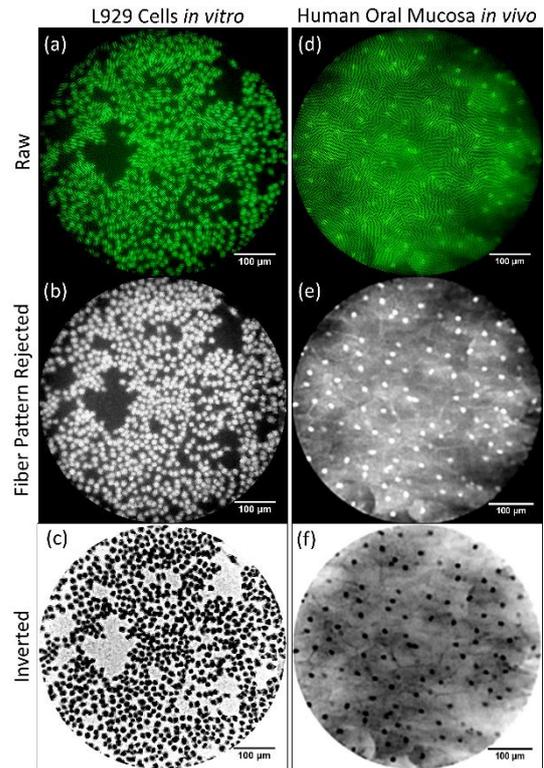

**Fig. 4** Raw (a), fiber pattern rejected (b) and inverted (c) images of a single layer of L929 cells *in vitro*. Raw (d), fiber pattern rejected (e) and inverted (f) images of normal human oral mucosa *in vivo*.

To test the feasibility of the microendoscope for imaging living cells, both L929 cell lines and oral mucosa from a volunteer (IRB review exempted) were imaged. Images were taken immediately after proflavine (0.01% wt/vol in PBS) was applied on the surfaces of the cell line or oral mucosa. The experiment setup to take images from a single layer of L929 cells in a disk was the same as that used for the *ex vivo* tissues. To image the oral mucosa the fiber bundle was handheld and brought in gentle contact with inner cheek of the volunteer. Fig. 4(a) and 4(b) show the raw and fiber pattern rejected fluorescence images collected from the L929 cells. The cells can be easily identified with little overlap. It is important to note that the bright spots in the images represent the nuclei instead of the cells because proflavine labels the cell DNA. Fig. 4(d) and 4(e) show the raw and fiber pattern rejected images collected from the oral mucosa. The nuclei of the mucosal cells can be clearly visualized with some background fluorescence which is attributed to the underlying cells and tissue scattering.

The images were further processed to extract quantitative information about the samples, such as cell density for cell lines and nucleus-to-cytoplasm ratio (N/C) for tissues. Quinn et al.[5] demonstrated that the use of N/C measured from 26 patients in Botswana has achieved a sensitivity of 86% and specificity of 87% in differentiating CIN2+ lesions from non-neoplastic cervical tissues. A median filter was used to reduce the outliers and bring out the core of the bright fluorescent dots. Then the images were inverted to further emphasize the dots. Finally, an 'unsharp mask' with a radius of 5 made the images even sharper, as shown in Fig. 4(c) and 4(f). From Fig. 4(c), it was determined, using a particle analysis function of ImageJ, that there were a total of 852 cells within the ROI. Thus the cell density was 2973 cells/mm$^2$, which is very close to the number estimated from a phase contrast microscope image (~3100 cells/mm$^2$). Similarly, the N/C of the oral mucosa in Fig. 4(f) was calculated to be 3.5%.

The results obtained with the experimental setup demonstrate the feasibility of using the smartphone microendoscope for high resolution fluorescence images. The image quality is comparable to that achieved with the high resolution microendoscopes (HRME) based on a scientific camera[21] or a DSLR camera[22]. However, the smartphone microendo-scope has a number of advantages over the HRME systems. Firstly, while scientific or DSLR cameras cost over $1,000 and very few people own one, smartphones, especially used smartphones, are widely available at low cost even in ruaral areas in LMIC. Being able to use customers' existing smartphones for imaging significantly increases the adoption of the technology in resource-poor settings. Secondly, the HRME uses a local computer to collect or download images and often requires a trained engineer on site to operate the system. The smartphone microendoscope does not need a local computer and the application software can be made easy to use, thus further reduces the cost associated with each use of the device in LMIC. More importantly, due to the convenient Internet access through a mobile data plan that is more widely available than a Wi-Fi network a smartphone microendoscope is more likely to be used as a point-of-care device for telemedicine applications. Finally, the experimental system measures about 20×15×5 cm (L×W×H) and weights only 612 grams. The final version of the smartphone microendoscope can be readily engineered to a handheld device.

The biggest challenges in implementing the smartphone microendoscope are: (1) the much smaller sensor pixel size of a smartphone camera than that used in the HRME systems and (2) the unchangeable built-in lens kit. Due to the low throughput of the imaging bundle it is critical to optimize the efficiency of the imaging optics so that a compararble signal-to-noise ratio can still be acieved with the smaller pixels of the smartphone cameras. An eyepiece has also been used with the objective to correctly image the fiber bundle on to the smartphone camera through the built-in lens kit. Although the microendoscope described in this report was specifically designed for proflavine as the contrast agent, it can be readily modified for applications with other contrast agents.

In conclusion, we have demonstrated the feasibility of a smartphone-based fiber optic microendoscope for high resolution fluorescence imaging. When used with proflavine the device can visualize cell nuclei in *ex vivo*, *in vitro* and *in vivo* biological samples. The technology provides a compact, lower cost, and 'smart' device which can potentially be used for early detection of neoplastic changes in various internal organs in LMIC.

The authors would like to acknowledge Dr. Rebecca K. Willits and Wafaa Nasir for providing the L929 cell lines.

*References*

1. WHO, "All cancers (excluding non-melanoma skin cancer) estimated incidence, mortality and prevalence worldwide in 2012," *World Health Organization.* http://globocan.iarc.fr/Pages/fact_sheets_cancer.aspx (2012).
2. S. Khondee et al., "Progress in molecular imaging in endoscopy and endomicroscopy for cancer imaging," *J Healthc Eng.* **4**(1), 1-22 (2013).
3. A. F. Gmitro et al., "Confocal microscopy through a fiber-optic imaging bundle," *Optics Letters.* **18**(8), 565-567 (1993).
4. G. Oh et al., "Optical fibers for high-resolution in vivo microendoscopic fluorescence imaging," *Optical Fiber Technology.* **19**(6 Part B), 760–771 (2013).
5. M. K. Quinn et al., "High-resolution microendoscopy for the detection of cervical neoplasia in low-resource settings," *PLoS One.* **7**(9), e44924 (2012).
6. T. Xie et al., "Fiber-optic-bundle-based optical coherence tomography," *Opt Lett.* **30**(14), 1803-1805 (2005).
7. T. J. Yoon et al., "Recent advances in photoacoustic endoscopy," *World J Gastrointest Endosc.* **5**(11), 534-539 (2013).
8. W. L. Curvers et al., "Endoscopic trimodal imaging versus standard video endoscopy for detection of early Barrett's neoplasia: a multicenter, randomized, crossover study in general practice," *Gastrointest Endosc.* **73**(2), 195-203 (2011).
9. N. D. Parikh et al., "In vivo diagnostic accuracy of high-resolution microendoscopy in differentiating neoplastic from non-neoplastic colorectal polyps: a prospective study," *Am J Gastroenterol.* **109**(1), 68-75 (2014).
10. ITU, "The world in 2014: ICT facts and figures," *The International Telecommunication Union.* http://www.itu.int/en/ITU-D/Statistics/Pages/facts/default.aspx (2014).
11. D. N. Breslauer et al., "Mobile phone based clinical microscopy for global health applications," *PLoS One.* **4**(7), e6320 (2009).
12. N. A. Switz et al., "Low-cost mobile phone microscopy with a reversed mobile phone camera lens," *PLoS One.* **9**(5), e95330 (2014).
13. D. Tseng et al., "Lensfree microscopy on a cellphone," *Lab Chip.* **10**(14), 1787-1792 (2010).
14. H. Zhu et al., "Cost-effective and compact wide-field fluorescent imaging on a cell-phone," *Lab Chip.* **11**(2), 315-322 (2011).
15. C. J. Wu et al., "An innovative smartphone-based otorhinoendoscope and its application in mobile health and teleotolaryngology," *J Med Internet Res.* **16**(3), e71 (2014).
16. P. J. M. Jongsma et al., "Device for coupling an endoscope to a videophone" *U.S.P. Application.* US 20060215013 A1 (2006)
17. A. Beery, "Introduction to mobileOCT's multimodal imaging," *MobileOCT.* http://www.mobileoct.com/introduction-to-multimodal.html (2014).
18. T. J. Muldoon et al., "Noninvasive imaging of oral neoplasia with a high-resolution fiber-optic microendoscope," *Head Neck.* **34**(3), 305-312 (2012).
19. T. J. Muldoon et al., "Evaluation of quantitative image analysis criteria for the high-resolution microendoscopic detection of neoplasia in Barrett's esophagus," *J Biomed Opt.* **15**(2), 026027 (2010).
20. M. Elter et al., "Physically motivated reconstruction of fiberscopic images," *ICPR 2006.* **3**, 599-602 (2006).
21. M. Pierce et al., "High-resolution fiber-optic microendoscopy for in situ cellular imaging," *J Vis Exp.* (47), (2011).
22. D. Shin et al., "A fiber-optic fluorescence microscope using a consumer-grade digital camera for in vivo cellular imaging," *PLoS ONE.* **5**(6), e11218 (2010).